# Polypeptide A9K at Nanoscale Carbon: Simulation Study

Vitaly V. Chaban, Andre Arruda and Eudes Eterno Fileti

Instituto de Ciência e Tecnologia, Universidade Federal de São Paulo, 12231-280, São José dos Campos, SP, Brazil

**Abstract**. An amphiphilic nature of the surfactant-like peptides is responsible for their propensity to aggregate at the nanoscale. These peptides can be readily used for a non-covalent functionalization of nanoparticles and macromolecules. This work reports an observation of supramolecular ensembles consisting of ultrashort carbon nanotubes (USCNTs), graphene (GR) and A9K polypeptide formed by lysine and arginine. Potential of mean force (PMF) is used as a major descriptor of the CNT-A9K and GR-A9K binding process, supplementing structural data. The phase space sampling is performed by multiple equilibrium molecular dynamics simulations with position restraints, where applicable. Binding in all cases was found to be thermodynamically favorable. Encapsulation in the (10,10) USCNT is particularly favorable. Curvature of external surface does not favor binding. Thus, binding of A9K at GR is stronger than its binding at the outer sidewall of USCNTs. Overall, the presented results favor non-covalent functionalization of nanoscale carbons that are considered interesting in the fields of biomaterials, biosensors, biomedical devices, and drug delivery.

**Introduction**

Thanks to an outstanding set of electrical, mechanical, thermal, and chemical properties, carbon nanotubes (CNTs) are widely investigated and utilized during the last two decades. CNTs are also considered promising for biomedicine and biotechnology.[1-3] Modification of CNTs by biological, organic and polymeric molecules and specific functional groups give rise to new classes of compounds. Many of these derivatives are still purely investigated.[4-7] Synthesis of CNTs has progressed appreciably during the last years. Several studies discuss well-controlled density, length, diameter, alignment of the sidewalls (vertically aligned CNTs), and number of sidewalls.[8, 9] A carbon nanotube with the length varying between 2-4 nm (ultrashort nanotube, USCNT) were obtained using density gradient separation method.[10, 11] This synthesis method enables a very narrow length distribution of the USCNT. In essence, a narrow USCNT is similar to lipids and hydrocarbons. The USCNT maintain some specific properties of longer CNTs including hydrophobicity.[11] Due to small length, it cannot be metallic. Hydrophobic nature of the all nanotubes leads to their aggregation in virtually all solvents. Little solubility of nanotubes in nonpolar solvents occurs due to a different size of solvent and solute molecules, rather than only due to a weak solvent-solute binding. Aggregation of CNTs and USCNTs may be both useful and harmful for applications in the biomedical fields.[12] For instance, the drug vector applications would benefit from USCNTs finely dispersed in an aqueous media.

We believe that covalent and non-covalent functionalizations are able to significantly modulate aggregation of USCNTs in solutions. Fluorination,[13] oxidation,[14] hydroxylation,[15] nitration[16] and grafting other groups was shown to take place readily with a satisfactory to good outcome of the main reaction product. Structure defects obviously matter. These reactions are conducted at a reasonable cost.[17] In turn, non-covalent functionalization is being actively developed. Non-covalent functionalization allows to preserve chemical and electrical properties of pristine nanotubes. Dispersion of CNTs of various length was achieved in surfactant molecules,[18, 19] ionic liquids,[20] polymers,[21] biological molecules, such as single-stranded DNA[22, 23] and peptides.[12]

Binding of biological materials with CNTs and particularly with USCNTs is interesting in the context of biosensors, toxicology research, and hybrids biomaterials. We expect this binding to increase solubility and dispersibility of nanotubes. In the meantime, desirable properties, such as an increased biocompatibility and therapeutic action, can be conferred. USCNTs have been already used as artificial nanopores inserted into a lipid bilayer. Ion transport and DNA translocation through these nanopores are different from previous studies with longer single-walled CNTs.[24] Luo and coworkers used NMR, CD, and ThT/fluorescence spectroscopy coupled with AFM imaging to study pH-dependent molecular interactions between CNTs and amyloid-beta peptide (Aβ). This peptide is considered to participate in the formation of Alzheimer's disease. CNTs affect self-organization of Aβ peptide chains engendering a different class of β-sheet-rich yet non-amyloid fibrils.[25] In the recent study, a short α-helical peptide was investigated as a model for encapsulation of biologically relevant macromolecules in single-walled CNTs using molecular dynamics (MD) simulations and free energy calculations.[26-29] Although geometry of CNT does not directly affect encapsulation of macromolecules, it significantly affects an equilibrium arrangement of the penetrated water molecules.[26] Raffaini and Ganazzoli investigated protein adsorption on CNTs and graphene (GR). According to these authors, proteins interact favorably with hydrophobic carbonaceous surfaces, irrespective of a secondary structure. Furthermore, proteins solubilize CNTs in water through an adsorption at their outer surfaces, as proven experimentally.[29]

Atomistic-precision computer simulations constitute a robust and sufficiently affordable tool to shed light on the interactions between macromolecules and nanoparticles. Provided that pairwise atom-atom interaction functions are parametrized adequately, simulations bring information concerning structure and thermodynamics of each simulated system. This information is, in many cases, inaccessible or accessible to quite a limited extent for the experimental studies due to technical reasons and resolution complications.

In this paper, we employ MD simulations to investigate structure properties and potential of mean force (PMF) between USCNTs and polypeptide A9K. Translational self-diffusion is discussed briefly in light of the A9K adsorption and encapsulation. The A9K polypeptide is formed by amino acids lysine and arginine. A9K was chosen for this study, because it exhibits an antibacterial action (medical effect) and amphiphilic nature (better solubilization). Also, usage of A9K to solubilize nanoscale carbonaceous structures has never been investigated before.

**Methodology**

**Simulated objects.** The systems for MD simulations investigated in this work are composed by the $A_9K$ amphiphilic peptide in pure water and $A_9K$ coupled to GR and USCNTs also in water. $A_9K$ is composed by nine alanine residues (hydrophobic moiety) and a charged lysine head group (hydrophilic moiety), as shown in Figure 1a. We consider (5,5), (10,10), and (20,20) USCNTs to understand an effect of surface curvature. An infinite GR sheet can be considered as a zero-curvature nanotube (Figure 1b). These simulations and these models do not account for electronic properties of GR.

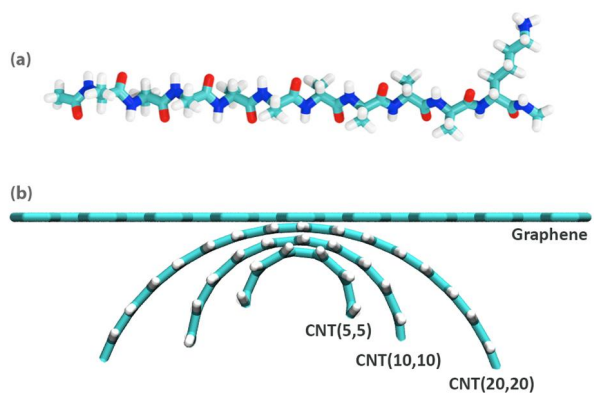

**Figure 1**. a) Molecular structure of the A9K peptide with N-terminal acetylation and C-terminal methylamidation: carbon atoms are cyan, hydrogen atoms are white, oxygen atoms are red, nitrogen atoms are blue. b) Molecular structure (in scale) of the nanotubes and graphene. Compare surface curvatures.

**Simulated systems.** Pairwise potential based all-atomistic MD simulations, based on the second Newton's law, were used to obtain structure properties of the A9K peptide in water and when coupled with USCNTs and GR (Figure 2). Seven independent MD simulations were performed: (1) lone solvated A9K (in water); (2) A9K inside the (10,10) USCNT; (3) A9K inside the (20,20) USCNT; (4), (5), (6) A9K outside the (5,5), (10,10), (20,20) USCNTs, respectively; (7) A9K and GR. Basic description of the simulated systems is provided in Table 1. The positive charge of the lysine residue was neutralized using the chloride anion. Keeping periodic systems electrostatically neutral is important for methodological considerations.

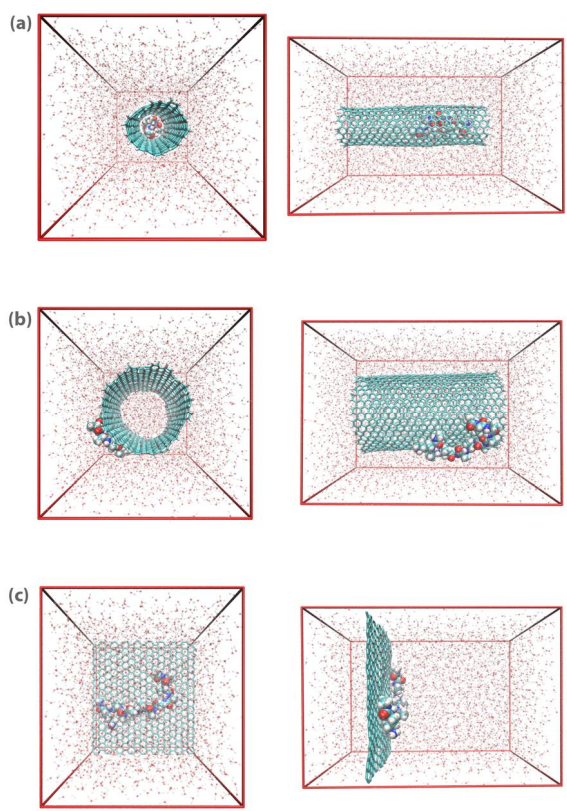

**Figure 2**. Front view (left) and lateral (right) of the equilibrated configurations of three investigated systems: (a) A9K inside the (10,10) USCNT; (b) A9K outside the (20,20) USCNT; (c) A9K at the graphene sheet.

Table 1. List of the simulated systems and their basic features

| # | Name of system | Location of peptide | N(C) in USCNT/GR | N (H$_2$O) | N$_{atoms}$ | Density of system, kg m$^{-3}$ |
|---|---|---|---|---|---|---|
| 1 | A9K | — | — | 3464 | 10527 | 997 |
| 2 | A9K+(10,10) | inside | 800 | 4329 | 13952 | 1017 |
| 3 | A9K+(20,20) | inside | 1600 | 5088 | 17069 | 1060 |
| 4 | A9K+(5,5) | outside | 400 | 3510 | 11075 | 1003 |
| 5 | A9K+(10,10) | outside | 800 | 4323 | 13934 | 1024 |
| 6 | A9K+(20,20) | outside | 1600 | 5094 | 17087 | 1064 |
| 7 | A9K+GR | — | 446 | 2831 | 9230 | 1018 |

**Models and algorithms**. The discussed MD simulations were performed using the well-known GROMACS MD simulation package,[30] version 5. Solvent (water) molecules were represented explicitly using the TIP3P model.[31] Carbon nanotubes, graphene sheet, and A9K peptide were represented using the CHARMM36 force field.[32] The leapfrog integration and propagation algorithm was applied to the equations-of-motion with a time-step of 2.0 fs. Increased time-step is possible thanks to constraining all carbon-hydrogen covalent bonds, whose vibration frequency is relatively high. The constant temperature, 310 K, was maintained by the Bussi-Donadio-Parrinello velocity rescaling thermostat[33] (with a relaxation time constant of 0.1 ps). This thermostat provides a correct velocity distribution for a given statistical mechanical ensemble. The constant pressure, 1.0 bar, was maintained by Parrinello-Rahman barostat[34] with a time constant of 1.0 ps and a compressibility constant of $4.5 \times 10^{-5}$ bar$^{-1}$. Every system sampled the phase space during 110 ns. The first 10 nanoseconds of simulations were regarded as equilibration. Proper equilibration of systems was proven by analysis multiple thermodynamic and structure components vs. time and averaged in blocks (0-2 ns, 2-4 ns ... 10-12 ns). Atomic configurations were saved every 10 ps generating a set of 5,000 configurations used in the subsequent statistical analysis. An average box size of all MD systems amounted to ca. 4.5 nm×4.5 nm×7.0 nm. Nanotubes were kept aligned with the principal axis of the MD box applying flat-bottom restraints with a harmonic force constant of 2000 kJ mol$^{-1}$ nm$^{-2}$. Periodic boundary conditions were used in all simulations to eliminate undesirable boundary/surface effects. Electrostatic interactions were calculated directly by Coulomb law up to 1.2 nm of separation between each two interaction sites. Electrostatic interactions beyond 1.2 nm were described by a computationally efficient implementation of Particle-Mesh-Ewald (PME) method.[35] Lennard-Jones-12-6 interactions were smoothly brought down to zero from 1.1 to 1.2 nm using the classical shifted force technique. This is was done for the sake of total energy conservation.

**Potential of mean force.** To obtain quantitative information regarding an interaction of the A9K peptide with the nanoscale carbons in water, potential of mean force (PMF) was recorded. PMF is a one-dimensional function that describes how the free energy between A9K and CNTs/GR evolves with an increase of distance between their centers-of-mass. To calculate PMF, we employed the umbrella sampling technique[36] with the weighted histogram analysis method (WHAM).[37] This technique requires restraints to be applied at specific positions along the reaction coordinate, so that all regions are appropriately sampled. In this particular case, 20 (systems # 4, 5, 6, 7) and 38 (systems # 2, 3) distances (with an increment of 0.1 nm) between centers-of-masses of the investigated molecules were restrained using a harmonic restraint with a force constant of 2000 kJ mol$^{-1}$ nm$^{-2}$. Note that the reaction coordinate coincided with different box vectors depending on the simulated MD system. The reaction coordinate evolved along the (0,0,1) vector in systems # 2, 3, 7. In all other systems, the reaction coordinate evolved along (1,1,0), see Figure 3. Each position of the reaction coordinate was sampled for 20 ns, whereas the first five nanoseconds of each such simulation were regarded as an equilibration.

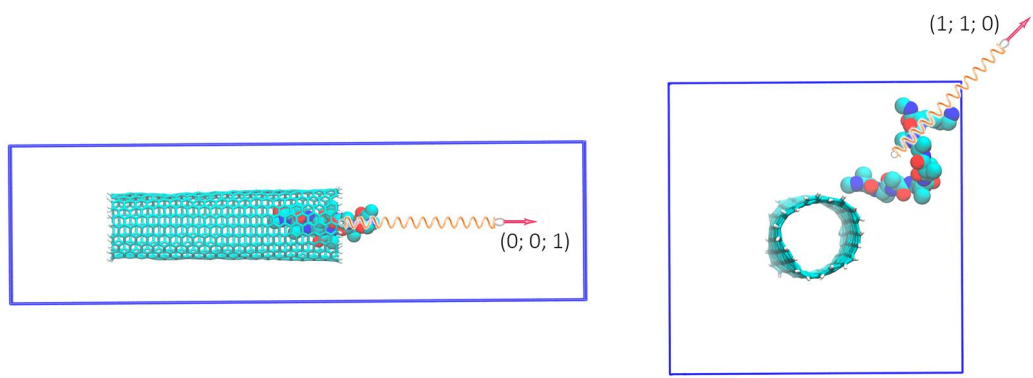

**Figure 3**. The pulling directions of the A9K peptide inside and outside USCNTs. The distance between the centers-of-masses of A9K and USCNTs was harmonically restrained.

**Results and discussions**

Interaction of the A9K peptide with USCNTs and GR decreases its solvent accessible area. Consequently, a time-averaged number of hydrogen bonds between A9K and water should decrease as well. Figure 4 shows radial distribution functions (RDFs) between the A9K carboxyl oxygen atoms and the water hydrogen atoms. These distributions correspond to prospective hydrogen bonds. The integrals of these RDFs are numerically equal to an average number of hydrogen bonds between the carboxyl group of A9K and water.

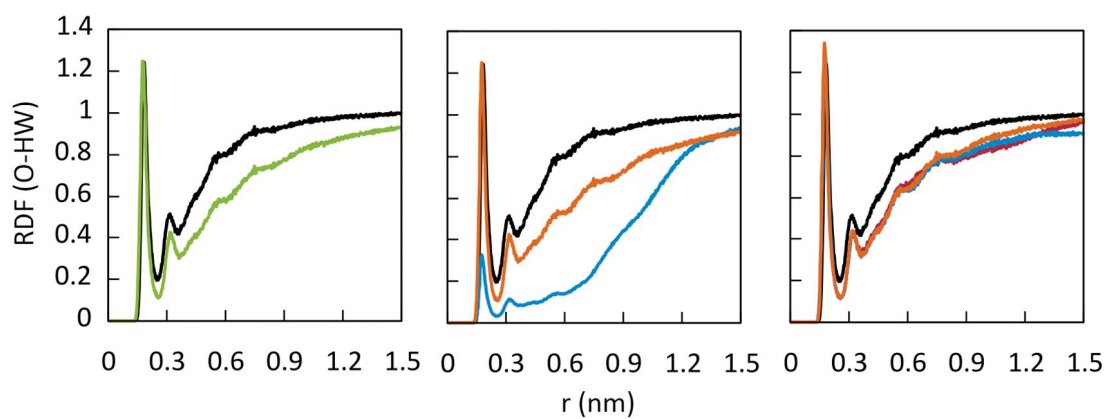

**Figure 4**. Radial distribution functions (RDFs) between A9K carboxyl oxygen atoms and water hydrogen atoms. RDFs in pure water are depicted by black curves in all plots. Left: at GR (green). Middle: A9K at the (10,10) and (20,20) USCNTs (blue and orange,

respectively), Right: A9K outside the (5,5), (10,10) and (20,20) USCNTs (red, blue and orange, respectively).

Due to a specific symmetry of USCNT, it may be insightful to plot structure distributions in two dimensions. Figure 5 depicts a radial distribution between the center-of-mass of A9K and the axis of USCNT. USCNT is hereby assumed to be an ideal cylinder. A9K tends to be adsorbed at the inner or outer surface of USCNT depending on its initial position. Therefore, the peptide at the nanotube surface corresponds to a local minimum potential energy state. Simulation of A9K penetration inside USCNTs or desorption from the inner cavity of USCNTs in real-time is not the scope of our investigation. Smaller curvature of USCNT favors better adsorption. Compare, maxima at 0.38 nm and 0.49 nm inside the (10,10) and (20,20) USCNTs, respectively. One would expect that adsorption of A9K at graphene (zero surface curvature) is even more energetically favorable. This issue will be numerically considered using potential of mean force calculations below.

In the case of A9K outside USCNTs (Figure 5), the center-of-mass of the peptide is separated by ca. 0.33 nm from the USCNT sidewall. The computed distance of 0.33 nm appears quite close to the van der Waals diameter of the carbon atom, 0.34 nm. Therefore, A9K is readily adsorbed at the outer surface of all nanotubes. It is an interesting result that A9K is able to efficiently wrap even the narrowest (5,5) USCNT. That is, A9K exhibits a high conformational flexibility to accommodate to different curvatures.

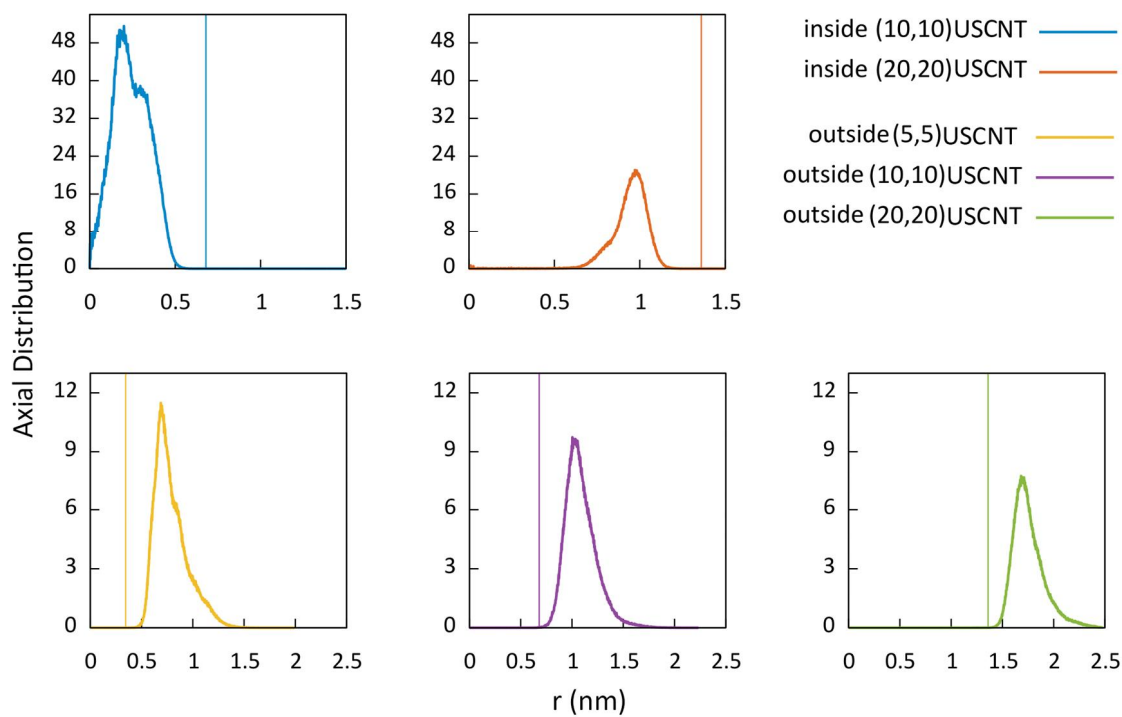

**Figure 5.** Radial distribution functions calculated in the XY-plane between the axis of the USCNT and the A9K peptide center-of-mass: inside the tube (top) and outside the tube (bottom). Vertical color lines stand for a position of the USCNT sidewall.

Table 2 summarizes structure and dynamics of the investigated systems. When adsorption occurs at the outer carbonaceous surface, radius of gyration, $R_g$, of the A9K peptide appears essentially unchanged in comparison with the fully hydrated case. Ranging between 0.8 and 0.9 nm, it remains well within error bars due to thermal motion. However, adsorption inside USCNTs is different: $R_g = 0.6$ nm inside the (10,10) USCNT and $R_g = 1.0$ nm inside the (20,20) USCNT. These alterations indicate substantial conformational changes of A9K due to confinement. Decreasing $R_g$ in the narrower tube is expected, since A9K must adjust its overall geometry to get encapsulated. In turn, increasing $R_g$ inside the (20,20) USCNT cannot be explained by the restricted volume conditions. Therefore, $R_g=1.0$ nm must be observed due to a significant adsorption energy, which exceeds energetic difference between those for various conformations of A9K.

Adsorption at the carbonaceous surfaces decreases a hydrogen bonding ability of the A9K peptide (Table 2). Whereas each A9K peptide generates 24 hydrogen bonds in water, this number ranges between 19 and 22 upon adsorption at the outer surface. Encapsulation in the USCNT brings even more drastic alterations: 19 H-bonds inside the (20,20) USCNT and 7 H-bonds inside the (10,10) USCNT. Most of these H-bonds are 0.21 nm long, as follows from Figure 4. Note that water molecules are able to penetrate the USCNT readily.[38]

Table 2. Average number of hydrogen bonds between A9K and water, radius of gyration $R_g$ (nm) and diffusion coefficient ($10^{-9}$ m$^2$ s$^{-1}$) of A9K peptide, D(A9K), and D(USCNT/GR)

| Systems | # HBs | $R_g$ | D(A9K) | D(USCNT/GR) |
|---|---|---|---|---|
| A9K in water | 24 ± 4 | 0.8 ± 0.2 | 0.7 ± 0.1 | — |
| A9K inside (10,10) USCNT | 7 ± 2 | 0.6 ± 0.0 | 0.7 ± 0.2 | 0.5 ± 0.1 |
| A9K inside (20,20) USCNT | 19 ± 2 | 1.0 ± 0.0 | 0.4 ± 0.2 | 0.1 ± 0.0 |
| A9K outside (5,5) USCNT | 20 ± 3 | 0.8 ± 0.1 | 0.4 ± 0.1 | 0.4 ± 0.1 |
| A9K outside (10,10) USCNT | 22 ± 2 | 0.8 ± 0.1 | 0.3 ± 0.1 | 0.3 ± 0.1 |
| A9K outside (20,20) USCNT | 21 ± 2 | 0.9 ± 0.1 | 0.3 ± 0.1 | 0.1 ± 0.0 |
| A9K on graphene | 19 ± 2 | 0.9 ± 0.1 | 0.0 ± 0.0 | 0.0 ± 0.0 |

Translational self-diffusion coefficients of A9K were calculated using mean-squared displacements (MSDs) of the A9K center-of-mass along the nanotube axis (Z). In the case of GR, MSDs were considered in the lateral (XY) direction. The slope of the MSD = f (time) curve over the time interval where this curve is a straight line gives a self-diffusion coefficient (D) numerically.[39] Diffusion coefficients of centers-of-mass of GR and USCNTs are provided for comparison. D inside the (10,10) USCNT is high due to a weak solvation (see the number of hydrogen bonds per peptide). Adsorption systematically decreases peptide diffusion, while the role of the surface curvature is not observed (Table 2). Diffusion of A9K at GR is zero. Thus, GR (zero-curvature carbonaceous surface) does not favor sliding of the adsorbed peptide. It is interesting to underline different Ds of the (10,10) USCNT in the systems with the same chemical composition, but different utilized adsorption sites: D=5×10$^{-10}$ m$^2$ s$^{-1}$ when A9K is inside USCNT and

D=3×10$^{-10}$ m$^2$ s$^{-1}$ when A9K is outside USCNT. This result indicates that adsorption of the A9K peptide is quite efficient as an interface between water and hydrophobic nanoscale carbons. Hydrophobic objects, e.g. fullerenes, diffuse very quickly, D=~1×10$^{-9}$ m$^2$ s$^{-1}$, in polar solvents.[40]

The stereochemistry of the peptide chain is best described by analyzing various orientations of two planar peptide units linked together by a central tetrahedral carbon atom. The torsional degrees of freedom of the N–C$_α$ (ϕ) and Cα–C′ (ψ) bonds permit the chain to sample conformational space. Several stable conformations corresponding to local energy minima are characterized by specific values of ϕ and ψ of each residue, Ramachandran plot. Figure 6 reports ϕ and ψ evaluated during the simulations for the ALA-3 and ALA-6 residues. ALA-3 is located nearly at the edge of the A9K peptide, while ALA-6 occupies an intermediate position. We note that the Ramachandran plot depicts local random and unfolded coil distributions of the both alanine residues in water, with large propabilities at β-strand and α-helix regions. In all cases of A9K except inside the (10,10) USCNT, the spatial arrangement of the involved amino acid residues corresponds to an unfolded structure. Inside the (10,10) USCNT, the confined peptide transitions to helical conformation, which is characterized by -115$^o$ < ϕ < -25$^o$ and -75$^o$ < ψ < -5$^o$. This significant structure alteration can be expected based on a small diameter of this nanotube. Radius of gyration of the confined A9K is hereby interpreted.

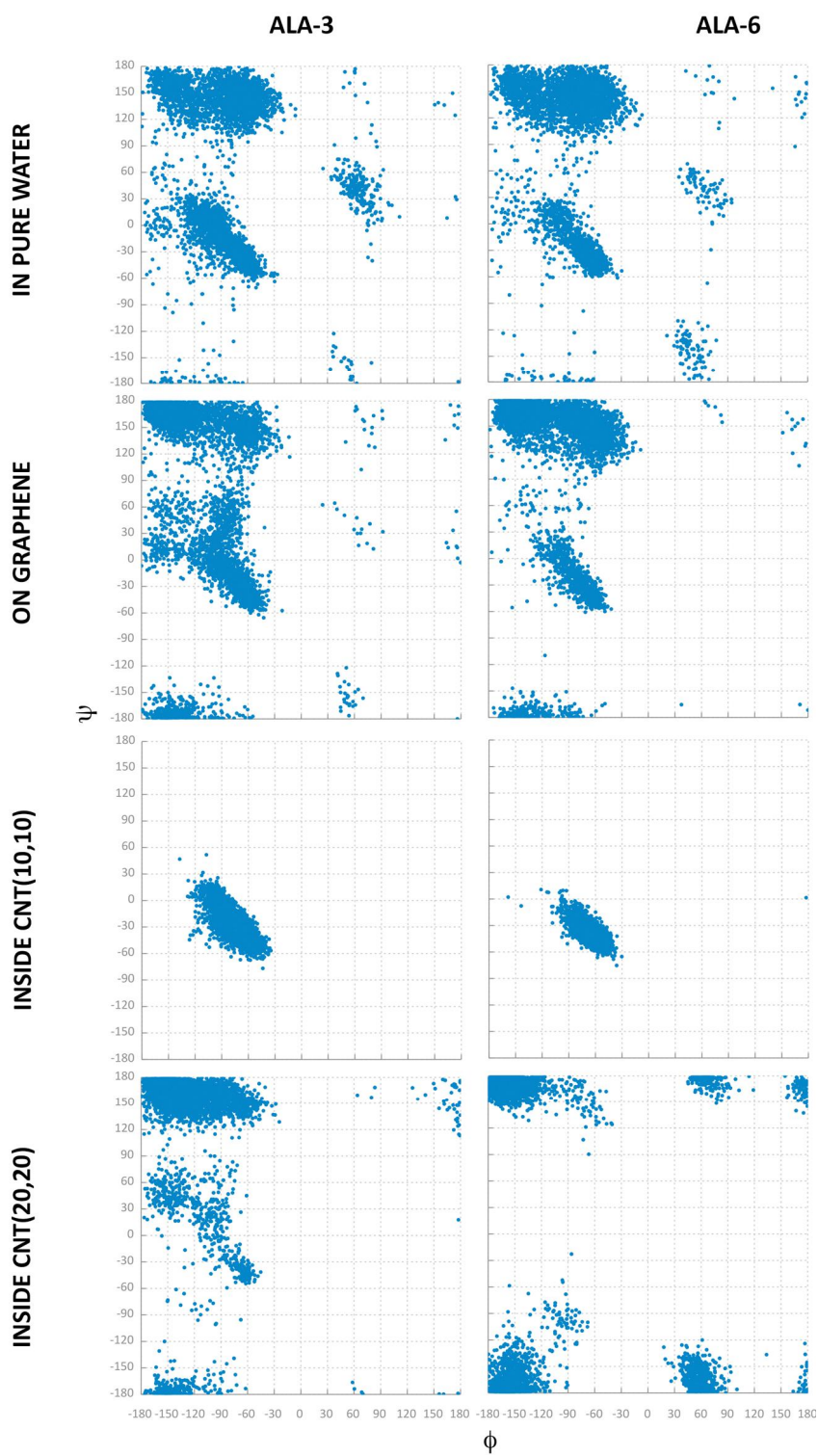

**Figure 6**. Ramachandran plots for two alanine residues for each simulated system. (Left) ALA-3 (peptide edge). (Right) ALA-6 (intermediate position).

Biomolecule non-covalent functionalization is a pathway to increase solubility of USCNTs in water. Furthermore, new properties can be conferred without changing inherent properties. This was experimentally demonstrated by using peptide and DNA binding to the nanotube sidewall.[41, 42] Thus, an interaction between carbonaceous surface and a given peptide is an important property for future chemical engineering. To explore thermodynamics of the A9K peptide adsorption at the internal and external surfaces of USCNTs and the surface of GR, we calculated PMFs (Figures 6-8) along the respective reaction coordinates, as introduced in the methodology.

Peptide─nanotube/graphene binding is of van der Waals nature. In turn, peptide-water binding is significantly of electrostatic nature. Nevertheless, equilibrium configurations of all the simulated systems involve the peptide adsorbed at the surface of USCNT/GR. These configurations constitute a starting state for the PMF calculation (the smallest separation distance, Figures 6-7). Adsorption of the A9K peptide at the outer carbonaceous surface of USCNT exhibits a weak dependence on the surface curvature. Indeed, the free energy gain due to binding ranges from 38 to 42 kJ mol$^{-1}$ in systems # 4, 5, 6. The observed difference is fairly small. When the A9K peptide is inside USCNT, curvature matters. For example, inside the (10,10) USCNT, A9K appears simultaneously adsorbed at the opposite inner surfaces, due to a small radius of cavity. The resulting free energy gain amounts to 217 kJ mol$^{-1}$, i.e. five times higher than inside the (20,20) USCNT, 42 kJ mol$^{-1}$ (Figure 8). Although binding of A9K to the (10,10) USCNT is still inferior to a single covalent bond, this free energy allows to claim a stable A9K@USCNT van der Waals complex. Van der Waals complex is a compound, which is maintained stable in the condensed phase thanks to predominant dispersion forces.

Graphene sheet is, in essence, a nanotube with zero surface curvature. The corresponding free energy gain due to A9K and GR binding is 91 kJ mol$^{-1}$, which falls between adsorption in the inner nanotube cavity and adsorption at the outer surface. Therefore, curvature of USCNT prevents adsorption of A9K, because the peptide has to somewhat adopt its conformation to that of the carbonaceous surface.

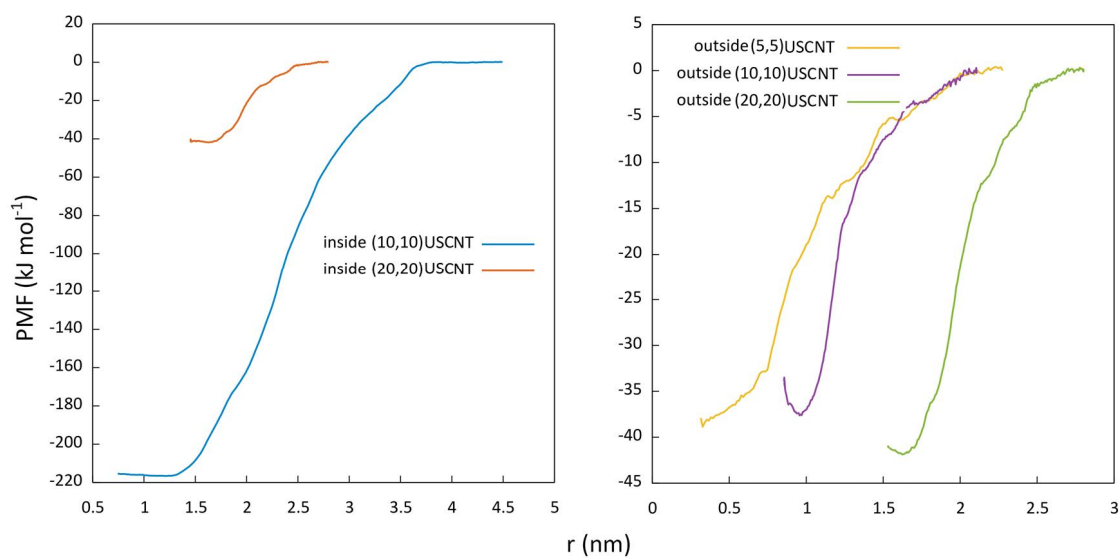

**Figure 6**. Potential of mean force upon separation of A9K from USCNTs in water. The depicted energies assume zero free energy at the largest considered separation. The distances were measured between center-of-mass of USCNT and center-of-mass of A9K, thus the distances are larger for the wider nanotubes.

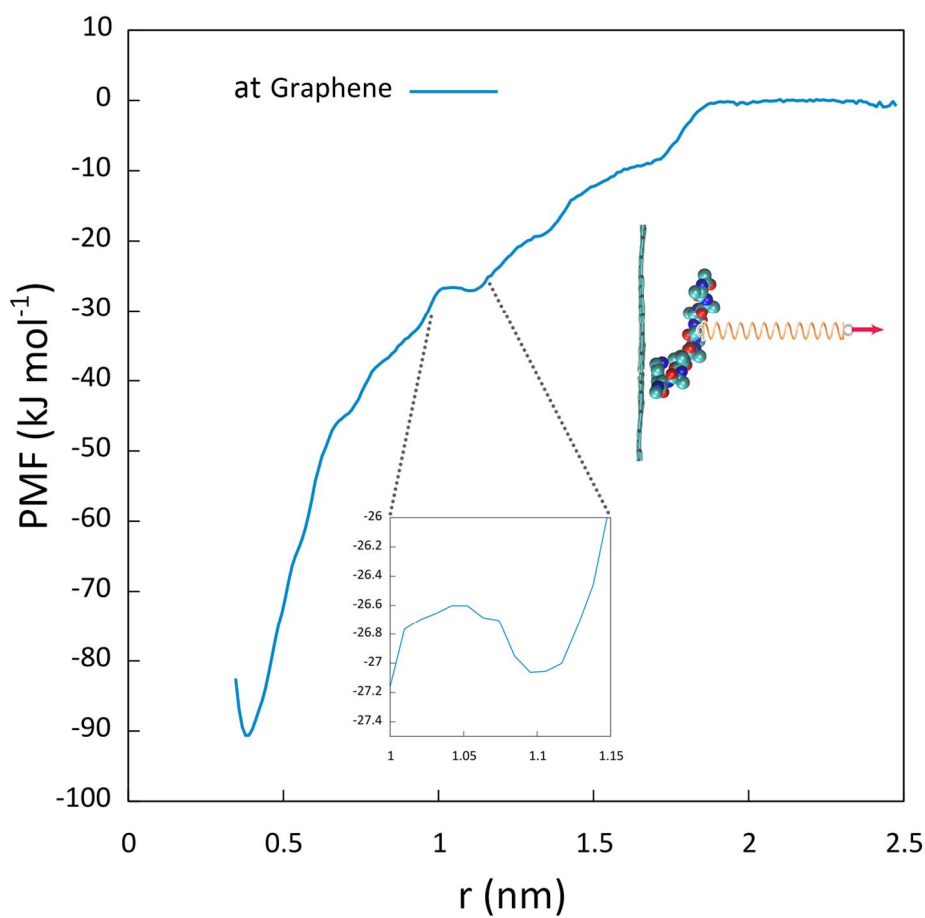

**Figure 7**. Potential of mean force upon separation of A9K from GR in water. The depicted energies assume zero free energy at the largest considered separation. The distances were measured between center-of-mass of USCNT and center-of-mass of A9K, thus the distances are larger for the wider nanotubes. Inset: local minimum related to the adhesion of the A9K edge to GR upon center-of-mass pulling.

According to Figure 8, binding of A9K with USCNT/GR is thermodynamically favorable, as compared to the state of this peptide in water. Local shallow minima, observed in Figures 6-7 at small separations, indicate partial connection of the A9K peptide to the respective surface. These local minima (see Figure 7, inset) occur at ca. 1.0 nm separation between the surface and the A9K center-of-mass, which is in line with the total length of the considered peptide.

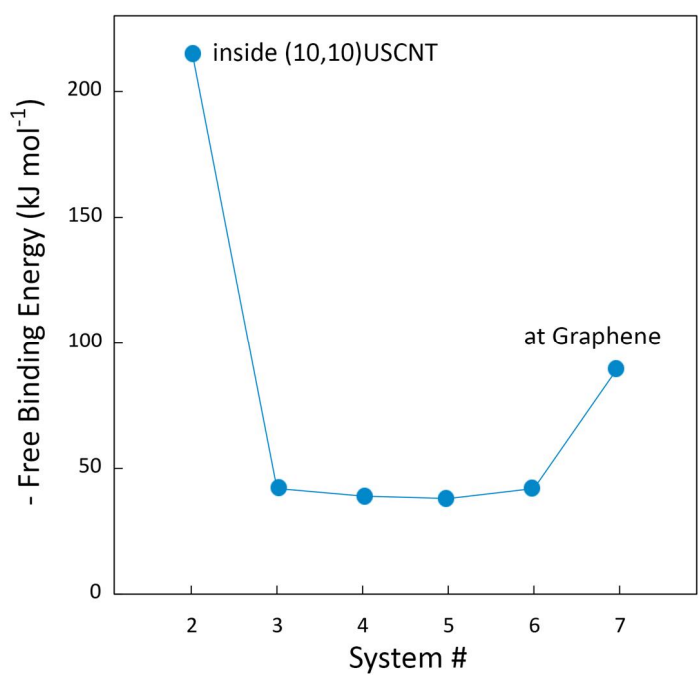

**Figure 8**. Free energy of binding between A9K and USCNTs/GR in different simulated systems. Free energy is the difference between the plateau region of the PMF curves (Figures 6-7) and maximum binding free energy. All attractive (i.e. negative) energies were multiplied by -1 for simplicity.

**Conclusions**

MD simulations of a few systems containing the A9K peptide, USCNT/GR and water have been conducted to unveil structure and thermodynamics. PMF calculations have been used to quantitatively describe binding between A9K and nanoscale carbon. We compare the (5,5), (10,10), and (20,20) USCNTs with the location of A9K inside and outside the tube. GR was used as a convenient model of USCNT with zero curvature. When A9K is confined inside the (10,10) USCNT, a transition from random coil to α-helix was observed. Adsorption and encapsulation of A9K at the carbonaceous surface is thermodynamically favorable. The free energy gain due to adsorption/encapsulation ranges from 40 kJ mol$^{-1}$ to 217 kJ mol$^{-1}$. It is an interesting finding that A9K readily penetrates the (10,10) USCNT, whose diameter is below 1.2 nm, assuming van der Waals diameter of the carbon atoms to equal to 0.34 nm.

Based on a reliable force field model, our results prove possibility of non-covalent functionalization of USCNTs and GR by the A9K peptide. Such functionalization is important to prepare dispersions of the nanoscale carbon in water and aqueous media, since solubility of pristine members of this family is negligible.

**REFERENCES**

1. N. Saito, H. Haniu, Y. Usui, K. Aoki, K. Hara, S. Takanashi, M. Shimizu, N. Narita, M. Okamoto, S. Kobayashi, H. Nomura, H. Kato, N. Nishimura, S. Taruta and M. Endo, *Chemical reviews*, 2014, DOI: 10.1021/cr400341h.
2. V. Castranova, P. A. Schulte and R. D. Zumwalde, *Accounts of chemical research*, 2013, 46, 642-649.
3. Y. Liu, Y. Zhao, B. Sun and C. Chen, *Accounts of chemical research*, 2013, 46, 702-713.
4. E. Heister, E. W. Brunner, G. R. Dieckmann, I. Jurewicz and A. B. Dalton, *ACS applied materials & interfaces*, 2013, 5, 1870-1891.
5. W. F. Chan, H. Y. Chen, A. Surapathi, M. G. Taylor, X. Shao, E. Marand and J. K. Johnson, *ACS nano*, 2013, 7, 5308-5319.
6. K. S. Siu, D. Chen, X. Zheng, X. Zhang, N. Johnston, Y. Liu, K. Yuan, J. Koropatnick, E. R. Gillies and W. P. Min, *Biomaterials*, 2014, 35, 3435-3442.
7. L. Wang, J. Shi, R. Liu, Y. Liu, J. Zhang, X. Yu, J. Gao, C. Zhang and Z. Zhang, *Nanoscale*, 2014, 6, 4642-4651.
8. G. A. Pilgrim, J. W. Leadbetter, F. Qiu, A. J. Siitonen, S. M. Pilgrim and T. D. Krauss, *Nano letters*, 2014, 14, 1728-1733.
9. F. Meng, R. Li, Q. Li, W. Lu and T.-W. Chou, *Carbon*, 2014, 72, 250-256.
10. X. Shi, B. Sitharaman, Q. P. Pham, F. Liang, K. Wu, W. Edward Billups, L. J. Wilson and A. G. Mikos, *Biomaterials*, 2007, 28, 4078-4090.
11. Y. Kuang, J. Liu and X. Sun, *The Journal of Physical Chemistry C*, 2012, 116, 24770-24776.

12. N. M. B. Cogan, C. J. Bowerman, L. J. Nogaj, B. L. Nilsson and T. D. Krauss, *The Journal of Physical Chemistry C*, 2014, 118, 5935-5944.
13. W. Zhang, P. Bonnet, M. Dubois, C. P. Ewels, K. Guérin, E. Petit, J.-Y. Mevellec, L. Vidal, D. A. Ivanov and A. Hamwi, *Chemistry of Materials*, 2012, 24, 1744-1751.
14. F. Morales-Lara, M. J. Pérez-Mendoza, D. Altmajer-Vaz, M. García-Román, M. Melguizo, F. J. López-Garzón and M. Domingo-García, *The Journal of Physical Chemistry C*, 2013, 117, 11647-11655.
15. Z. Liu, Y. Liu and D. Peng, *Journal of materials science. Materials in medicine*, 2014, 25, 1033-1044.
16. Y. Wang, S. V. Malhotra, F. J. Owens and Z. Iqbal, *Chemical Physics Letters*, 2005, 407, 68-72.
17. N. Arnaiz, M. F. Gomez-Rico, I. Martin Gullon and R. Font, *Industrial & Engineering Chemistry Research*, 2013, 52, 14847-14854.
18. N. Kocharova, J. Leiro, J. Lukkari, M. Heinonen, T. Skala, F. Sutara, M. Skoda and M. Vondracek, *Langmuir : the ACS journal of surfaces and colloids*, 2008, 24, 3235-3243.
19. M. F. Islam, E. Rojas, D. M. Bergey, A. T. Johnson and A. G. Yodh, *Nano letters*, 2003, 3, 269-273.
20. Y. Shim and H. J. Kim, *ACS nano*, 2009, 3, 1693-1702.
21. R. Guo, Z. Tan, K. Xu and L.-T. Yan, *ACS Macro Letters*, 2012, 1, 977-981.
22. S. Alidori, K. Asqiriba, P. Londero, M. Bergkvist, M. Leona, D. A. Scheinberg and M. R. McDevitt, *The journal of physical chemistry. C, Nanomaterials and interfaces*, 2013, 117, 5982-5992.
23. D. Roxbury, A. Jagota and J. Mittal, *The journal of physical chemistry. B*, 2013, 117, 132-140.
24. L. Liu, C. Yang, K. Zhao, J. Li and H. C. Wu, *Nat Commun*, 2013, 4, 2989.
25. J. Luo, S. K. Warmlander, C. H. Yu, K. Muhammad, A. Graslund and J. Pieter Abrahams, *Nanoscale*, 2014, 6, 6720-6726.
26. Z. S. Zhang, Y. Kang, L. J. Liang, Y. C. Liu, T. Wu and Q. Wang, *Biomaterials*, 2014, 35, 1771-1778.
27. Y. Kang, Q. Wang, Y. C. Liu, J. W. Shen and T. Wu, *The journal of physical chemistry. B*, 2010, 114, 2869-2875.
28. Y. Kang, Y. C. Liu, Q. Wang, J. W. Shen, T. Wu and W. J. Guan, *Biomaterials*, 2009, 30, 2807-2815.
29. G. Raffaini and F. Ganazzoli, *Langmuir : the ACS journal of surfaces and colloids*, 2013, 29, 4883-4893.
30. B. Hess, C. Kutzner, D. van der Spoel and E. Lindahl, *J. Chem. Theory Comput.*, 2008, 4, 435.
31. W. L. Jorgensen, J. Chandrasekhar, J. D. Madura, R. W. Impey and M. L. Klein, *J. Chem. Phys.*, 1983, 79, 926-935.
32. R. B. Best, X. Zhu, J. Shim, P. E. Lopes, J. Mittal, M. Feig and A. D. Mackerell, Jr., *J Chem Theory Comput*, 2012, 8, 3257-3273.
33. G. Bussi, D. Donadio and M. Parrinello, *J. Chem. Phys. 126*, 2007, 126, 014101-014108.
34. M. Parrinello and A. Rahman, *J. Appl. Phys.*, 1981, 52, 7182-7192.
35. T. Darden, D. York and L. Pedersen, *J. Chem. Phys.*, 1993, 98, 10089-10099.
36. G. M. Torrie and J. P. Valleau, *J. Comp. Phys.*, 1977, 23, 187.
37. S. Kumar, D. Bouzida, R. H. Swendsen, P. A. Kollman and J. M. Rosenberg, *J. Comp. Chem.*, 1992, 13, 1011.
38. V. Chaban, *Chemical Physics Letters*, 2010, 500, 35-40.

39. P. F. F. Almeida and W. L. C. Vaz, eds., *Lateral Diffusion in Membranes*, Elsevier Science, New York, 1995.
40. V. V. Chaban, C. Maciel and E. E. Fileti, *The journal of physical chemistry. B*, 2014, 118, 3378-3384.
41. S. Daniel, T. P. Rao, K. S. Rao, S. U. Rani, G. R. K. Naidu, H.-Y. Lee and T. Kawai, *Sensors and Actuators B: Chemical*, 2007, 122, 672-682.
42. Z. Kuang, S. N. Kim, W. J. Crookes-Goodson, B. L. Farmer and R. R. Naik, *ACS nano*, 2010, 4, 452-458.